\documentclass[11pt]{article}

\usepackage{graphics}   
\usepackage{endnotes}            
\usepackage{amsfonts}            
\usepackage{amssymb}             
\usepackage{amsthm}              
\usepackage{amsmath}            
\usepackage{algorithm2e}           

\usepackage[letterpaper]{geometry}
\usepackage{rotating}            

\usepackage{setspace}                
\usepackage{natbib}
\bibpunct{(}{)}{;}{a}{,}{;}

\begin{document}

\title{
Phase Transitions in the Edge/Concurrent Vertex Model\thanks{This work was supported by NSF awards DMS-1361425 and SES-1826589.}
}

\author{
Carter T. Butts\thanks{Departments of Sociology, Statistics, Computer Science, and Electrical Engineering and Computer Science; University of California, Irvine; buttsc@uci.edu}
}
\date{1/4/20}
\maketitle

\begin{abstract}
Although it is well-known that some exponential family random graph model (ERGM) families exhibit phase transitions (in which small parameter changes lead to qualitative changes in graph structure), the behavior of other models is still poorly understood.  Recently, Krivitsky and Morris have reported a previously unobserved phase transition in the edge/concurrent vertex family (a simple starting point for models of sexual contact networks).  Here, we examine this phase transition, showing it to be a first order transition with respect to an order parameter associated with the fraction of concurrent vertices.  This transition stems from weak cooperativity in the recruitment of vertices to the concurrent phase, which may not be a desirable property in some applications.\\[5pt]
\end{abstract}

\theoremstyle{plain}                        
\newtheorem{axiom}{Axiom}
\newtheorem{lemma}{Lemma}
\newtheorem{theorem}{Theorem}
\newtheorem{corollary}{Corollary}

\theoremstyle{definition}                 
\newtheorem{definition}{Definition}
\newtheorem{hypothesis}{Hypothesis}
\newtheorem{conjecture}{Conjecture}
\newtheorem{example}{Example}

\theoremstyle{remark}                    
\newtheorem{remark}{Remark}

\section{Introduction}

Krivitsky and Morris (personal communication) have recently demonstrated the presence of previously unobserved phase transition behavior in the edge/concurrent vertex model.  This is an interesting case, in that the transition differs from the well-known ``density explosion'' transitions \citep{strauss:siam:1986,handcock:ch:2003,butts:sm:2011b,schweinberger:jasa:2011} observed in the case of unbounded change statistic models (e.g., models containing subgraph census statistics whose highest-order statistic has a positive coefficient).  Here, we provide some theoretical analysis of the behavior of this model, focusing on the regime in which sparsity is favored and concurrency is suppressed (the most typical use case).  This analysis suggests that the model undergoes a first order phase transition with respect to an order parameter given by the fraction of non-concurrent nodes.  We suggest that this behavior is the result of a weak form of cooperativity, in which vertices ``recruited'' to the concurrent set become freed to ``recruit'' others, leading to a rapid emptying out of the non-concurrent set.  It should not be assumed that this behavior is necessarily ``unsocial,'' though it is unclear that it is realistic for e.g. sexual contact networks.

\subsection{Model}

The edge/concurrent vertex model is defined as follows.  Let $Y$ be a random graph with support on the set $\mathcal{Y}_N$ of order-$N$ simple graphs, and define $t_e$ and $t_c$ to be functions counting (respectively) the number of edges and the number of concurrent vertices (i.e., vertices with degree greater than 1) of their arguments.  The edge/concurrent vertex model is then
\begin{equation}
\Pr(Y=y|\theta) = \frac{\exp(\theta^T t(y)}{\sum_{y'\in\mathcal{Y}_N} \exp(\theta^Tt(y))} \mathbb{I}_{\mathcal{Y}_N} (y) \label{eq_ergm}
\end{equation}
where $\theta=(\theta_e,\theta_c)$ is a real parameter vector, $t=(t_e,t_c)$, and $\mathbb{I}_{\mathcal{Y}_N}$ is an indicator for membership in the support.  

This model should not be confused with the edge/concurrent tie model studied e.g. by \citet{butts:jms:2016}, which is itself equivalent to the edge/isolate model.  In that model, concurrency is enhanced or suppressed for every edge after the first that is incident on any given vertex.  Here, we see from the conditional representation
\begin{equation}
\log \frac{\Pr(Y=y^+_{ij}|\theta)}{\Pr(Y=y^-_{ij}|\theta)}  = \theta_e + \theta_c \left[I(d_i(y^-_{ij})=1)+I(d_j(y^-_{ij})=1)\right] \label{eq_condlogit}
\end{equation}
(with $I$ being an indicator of its argument, $d_k$ being the degree of $k$ in its argument, and $y^+_{ij}$,$y^-_{ij}$ being respectively the graph $y$ with the $i,j$ edge forced to be present or absent) that the edge/concurrent vertex model affects only the propensity to add edges to pendants (respectively, to remove them from vertices of degree two).  This seemingly minor difference is ultimately consequential, as we shall see.

Here, we focus on the regime in which (1) sparsity is favored and (2) concurrency is suppressed.  These conditions are obtained when $\theta_e<0, \theta_c<0$.  This is the most common use case, particularly in the context of sexual contact networks where a negative concurrency parameter may be used to parameterize norms discouraging multiple partnerships \citep{morris.et.al:ajph:2009}.\footnote{It should be noted that the case of $\theta_c>0$, \emph{pendant avoidance,} is also interesting in contexts for which relying on a single partner is penalized.  The most obvious example would be negative exchange networks \citep{willer:bk:1999}, in which pendant nodes are vulnerable to being placed in Bertrand competition.}

\section{Bernoulli Bounds Analysis}

It is natural to begin with a simple analysis of the system behavior using Bernoulli graph bounds \citep{butts:sm:2011b}.  To do so, we must for each $i,j$ edge variable identify the maximum and the minimum values of $\theta^T \Delta_{ij}(t,y)$ over all $y\in \mathcal{Y}_N$, where $\Delta_{ij}(t,y)= t(y^{+}_{ij})-t(y^{-}_{ij})$ is the \emph{changescore functional} evaluated at $y$.  Since this model is homogeneous, these bounds are the same for any choice of $i,j$, and we can see immediately from Equation~\ref{eq_condlogit} that our marginal edge probabilities are bounded by
\begin{equation}
\left[1+\exp(-\theta_e-2\theta_c)\right]^{-1} \le \Pr(Y_{ij}=1) \le \left[1+\exp(-\theta_e)\right]^{-1} \label{eq_bern_bounds}
\end{equation}
in the regime for which $\theta_c<0$.  For the sparse case in which $\theta_e\ll 0$, we can see that the lower bound is smaller than the upper bound by a factor of approximately $\exp(2 \theta_c)$.

Since $Y$ is bounded above by a homogeneous Bernoulli graph with expected density $\mathrm{logit}^{-1}(\theta_e)$ and below by a homogeneous Bernoulli graph with expected density $\mathrm{logit}^{-1}(\theta_e+2\theta_c)$, we see immediately that some forms of extreme behavior are impossible.  For instance, the model cannot undergo a ``density explosion'' in $N$ (since the expected density is bounded above by a constant), and indeed the model cannot exhibit degeneracies based on fixation of edge variables (i.e., the entropy of each edge variable cannot go to zero in $N$, since the changescores are strictly bounded).  More broadly, the insensitivity of the bounding graphs to $N$ suggests that the behavior of this model family is necessarily at least somewhat size stable.

At first blush, this analysis would seem to suggest that the edge/concurrent vertex model cannot exhibit complex phase behavior.  Certainly, it rules out many types of transitions common to badly specified ERGMs.  However, the gap between the bounds of Equation~\ref{eq_bern_bounds} still allows room for non-trivial phenomena, especially when $\theta_c \ll \theta_e$.  We now turn to a deeper analysis of this question.

\section{Phase Analysis}

Bernoulli bounds provide a very simple tool for ruling out certain types of behavior, but in-depth analysis of phase behavior requires more complex tools.  Here, we borrow techniques from statistical mechanics to characterize the phase transition of the edge/concurrent vertex model.

\subsection{Phase Characterization and Order Parameter}

\begin{figure}
\begin{center}
\includegraphics[width=0.7\textwidth]{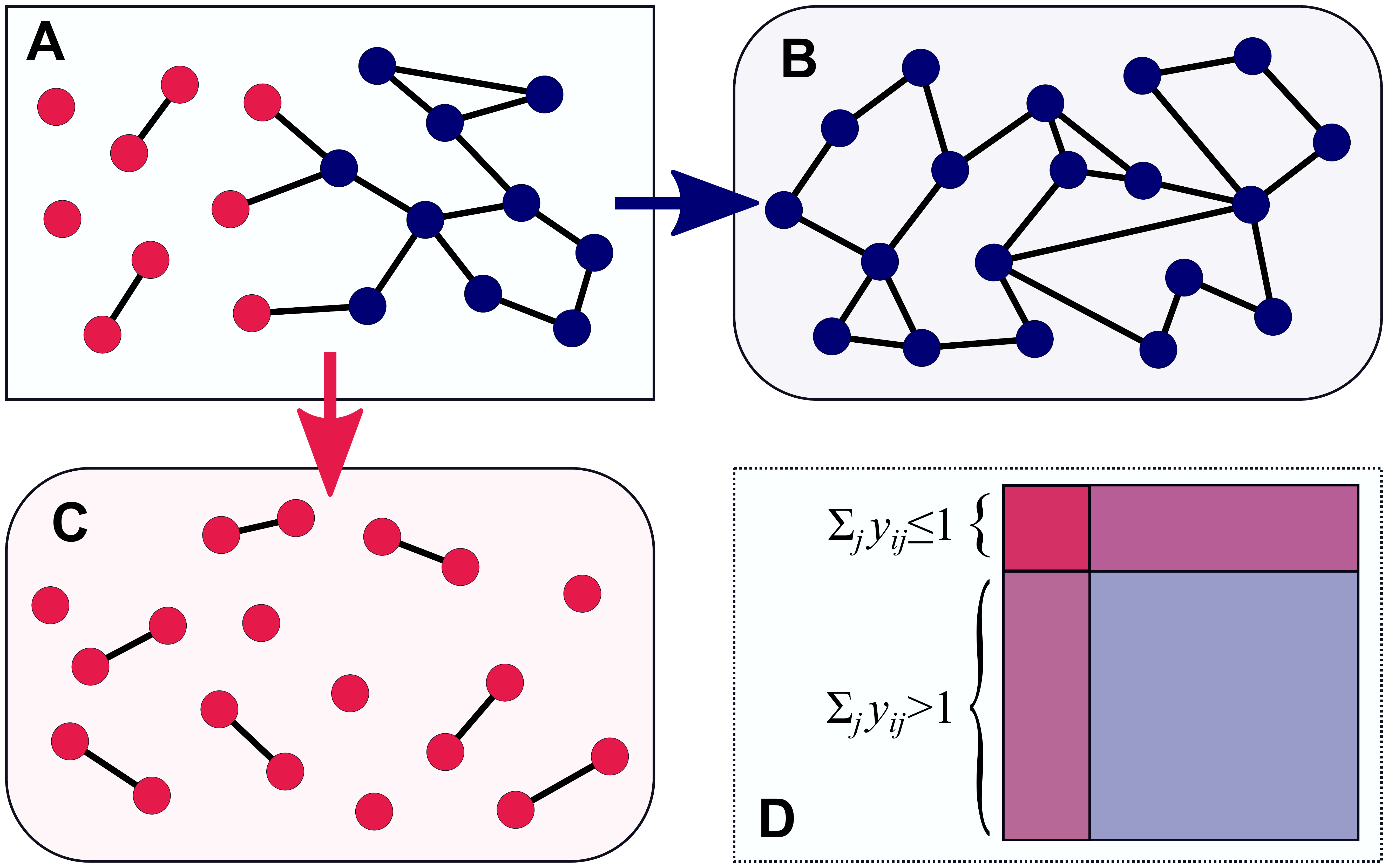}
\caption{ Phase behavior of the edge/concurrent vertex model.  (A) We characterize the graph in terms of concurrent (blue) and non-concurrent (red) phases, which may co-occur for some values of $\theta$ and $N$.  In the ``dense'' regime (B), concurrent vertices dominate (though overall density may remain fairly low); alternately, the model can also exhibit an ``ultra-sparse'' regime (C) in which almost all nodes are non-concurrent.  The transition between regimes can be extremely sharp.  (D) We divide the adjacency structure of the graph by phase when approximating the partition function.  The fraction of vertices in the sparse phase is used as an order parameter.  \label{fig_phases}}
\end{center}
\end{figure}

To proceed, we must first characterize the types of structures that make up the distinct phases of interest in graphs drawn from the edge/concurrent vertex model.  Here, we follow an approach similar to that used by \citet{butts:jms:2016} to characterize the edge/concurrent tie model, in separating the vertex set into a ``dense'' (or concurrent) phase containing all vertices of degree $\ge 2$, and a ``sparse'' (or non-concurrent) phase containing all vertices of degree $<2$ (see Figure~\ref{fig_phases}).  These phases may coexist---i.e., the same graph may contain vertices of both types---but within certain $\theta,N$ regimes one or another phase may strongly dominate.  In particular, Krivitsky and Morris report a sharp transition from an ``ultra-sparse'' regime with little concurrency to a (relatively) ``dense'' regime in which all or almost all vertices are concurrent.\footnote{This should not be confused with the ultra-dense regimes that arise e.g. from the edge-triangle model.  Where $\theta_e\ll 0$, these graphs may be quite sparse in an absolute sense.  They are, however, dense enough to have mean degree near or exceeding 2.}  Our focus is on the analysis of this transition.

To describe the phase behavior of the edge/concurrent vertex model, we require an \emph{order parameter} that characterizes the system state relative to the phases of interest.  An order parameter must take a value of 0 in the infinite-temperature limit (see below), taking on non-zero values as one moves from the maximally mixed to well-ordered states.  Here, a natural choice for the order parameter is the fraction of all vertices in the sparse phase (i.e., $m(y)=1-t_c(y)/N$).  This approaches 0 for large graphs in the high-temperature limit, and approaches 1 in the regime for which all vertices are non-concurrent.  Conveniently, this parameter is also an affine transformation of $t_c$, simplifying analysis.

\subsection{Physical Expression of the Edge/Concurrent Vertex Model}

For purposes of analysis, it is convenient to modify the expression of the edge/concurrent vertex model given in equation~\ref{eq_ergm} to a more physical form \citep[see e.g. ][]{grazioli.et.al:jpcB:2019}.  Specifically, we define a new parameter vector $\phi=(\phi_e,\phi_c)$ such that $-\phi^T t(y)/(k_B T) = \theta^T t(y)$, where $k_B$ is Boltzmann's constant and $T$ is a quantity we will here call the \emph{edge temperature.}  Since we have no external referents for temperature or energy, we are free to choose our units in a manner that is analytically convenient.  We begin by defining our energy unit as the baseline energy associated with a single edge (i.e., considering no other effects), which has the consequence of fixing $\phi_e=1$.  We now select our temperature scale such that $k_B=1$ in these units.  An immediate consequence of these choices is that the temperature of a model with parameter $\theta_e$ is equal to $-1/\theta_e$.\footnote{Note that it is possible for temperature to be negative; this is a common feature of systems for which the maximum energy per degree of freedom is bounded, and is not peculiar to ERGMs.}  (This in turn justifies our definition of $T$ as the ``edge temperature,'' since it is the value whose negative reciprocal is equal to the standard edge parameter.)  The quantity $\phi^T t(y)$ is analogous to the internal energy of $y$ (in ``edge units''), with lower-energy graphs being ceteris paribus more probable at positive temperature.  We observe in passing that the (tacit) choice of the counting measure in equation~\ref{eq_ergm} amounts to the assumption that each unique graph in $\mathcal{Y}_N$ constitutes a distinct microstate with unit entropy; when $Y$ is taken to arise from dynamics on unmodeled underlying degrees of freedom \citep[see e.g.][]{butts:jms:2019} this may be less than ideal.  In particular, note that entropic effects do not scale with temperature in the same manner as energetic ones, and hence ``offsets'' arising from entropic terms should \emph{not} be folded into $\theta$ for purposes of e.g. finding the critical temperature.  Here, we assume such effects to be absent.

As others have noted \citep[see e.g.][]{robins.et.al:ajs:2005}, high temperatures correspond to low-magnitude values of $\theta$; in the limit of infinite temperature, $\theta \to 0$ (yielding, in the case of the counting measure, a uniform random graph on the support).  Similarly, increasing $||\theta||$ amounts to cooling the system, ultimately concentrating probability mass on some ground state.  In a more conventional social interpretation, $T$ scales the effective social forces that drive structure formation: when $T$ is high, these forces are weak relative to idiosyncratic background factors, leading to highly randomized structures; when $T$ is low, these forces are relatively strong, driving the network to a rigidly selected conformation.  In the special case of utility-based dynamics \citep[e.g.][]{snijders:sm:2001,young:bk:1998,butts:sm:2007b}, the minima of the low-temperature limit correspond (under certain regularity conditions \citep{butts:pres:2009}) to Nash equilibria of the associated network game, obtained when actors are pure utility maximizers.  In that setting, $T$ may be very loosely interpreted as indifference, propensity to err in making decisions, or the scale of random ``circumstantial'' factors that whimsically alter actors' preferences in specific cases.  $T$ is not therefore without substantive meaning in non-physical settings, though our use of it does not depend on any particular substantive interpretation.

For purposes of the analyses below, it will be convenient to work with the partition function, $Z(\theta) = \sum_{y'\in \mathcal{Y}_n} \exp\left(\theta^T t(y')\right)$.  Since we will be interested in the variation in $Z$ (and related quantities in terms of temperature, it is often useful to employ the form $Z(T,\phi) = \sum_{y'\in \mathcal{Y}_n} \exp\left(-\phi^T t(y')/(k_b T)\right)$ instead. Likewise, we will need to consider partition functions associated with subsets of $\mathcal{Y}_N$.  In particular, we will work with $Z(T,\phi|M) = \sum_{y'\in \mathcal{Y}_n: m(y')=M} \exp\left(-\phi^T t(y')/(k_b T)\right)$, which is the partition function on the set of graphs with order parameter value $M$.  (Intuitively, this is simply the normalizing factor for the distribution of equation~\ref{eq_ergm} conditional on the order parameter.)  Finally, we define the \emph{free energy} of $Y$ to be $F_\phi(T)=-k_B T \log Z(T,\phi)$, with $F_\phi(T|M)=-k_B T \log Z(T,\phi|M)$ the free energy of the conditional distribution of $Y$ given the order parameter.  Physically, the free energy is a quantity that tends to be minimized by a system in energetic exchange with its environment at constant temperature\footnote{Or better, if we observe such a system at a random time, we are more likely to find it in low energy states than high energy states.}.  Statistically, the free energy is simply the negative log state probability, rescaled by the amount of energy needed to raise the entropy of the system by a fixed amount (one $log_e$ unit, or ``nat'').  In either interpretation, it is useful for characterizing model behavior as parameters are rescaled.

\subsection{Approximating the Partition Function}

An evident difficulty with the direct use of $Z$ is its well-known intractability.  While no closed form exists for the partition function of the edge/concurrent vertex model, we here introduce an approximation that can be computed in polynomial time.  To calculate $Z$, we must sum the numerator of equation~\ref{eq_ergm} over all microstates (possibly conditional on $t_c$); while this is an intractably large sum, we can exploit the homogeneity of the model family to work instead with \emph{states} defined by unique values of the sufficient statistics (of which there are many fewer), provided that we know the multiplicity of those states.  For the latter, we can leverage various related graph theoretic results to obtain exact counts or, in some cases, excellent approximations.

In developing our approximation, it is helpful to again think of the vertices of $Y$ as being divided into two sets (the sparse phase and the dense phase) based on degree (Figure~\ref{fig_phases}D).  Given an assignment of vertices to the respective phases, the number of possible graphs can then be approximated in terms of the number of graphs on the dense phase vertices having minimum degree two times the number of ways to assign edges into either the sparse phase or the interface between phases such that no sparse phase vertex has more than one edge.  (Note that our approximation arises in assuming that no dense phase vertex is dependent upon interfacial ties in order to meet its degree quota.  This assumption will be very good except in cases where the dense phase is very sparse and the sparse phase is large and (relatively) dense.  This seems to work well in the present case.) In practice, we will also ration the edges available to each set (since this is necessary to fix the state), summing numbers of possible graphs across edge values.

We now discuss each portion of the multiplicity calculation, before describing how they are combined to form the final partition function approximation.

\subsubsection{Multiplicity of the Sparse Phase (and Interface)}

Vertices in the sparse phase are characterized by having degree $\le 1$; these ties may be either to other sparse phase vertices, or to vertices in the dense phase (the latter of which may have any number of ties).  Let us assume that there are $n_s$ vertices in the sparse class, and $N-n_s$ vertices in the dense class.  Now, consider adding edges to either the sparse-class induced subgraph, or the sparse/dense edge cut.  Each edge placed between two sparse vertices ``consumes'' both (in the sense of making them unavailable to place further edges), while an edge placed in the cut consumes only the one sparse phase vertex.  It then follows that the maximum number of edges that may be placed is $\lfloor(n_s-\min\{n_s,N-n_s\})/2\rfloor + \min\{n_s,N-n_s\}$.  Now, assume that $E$ edges are to be placed among vertices within the sparse set.  The number of ways to do this without placing two edges on the same vertex is
\begin{equation}
C_{s}(E,n_s) = E! 2^E \binom{\lfloor n_s\rfloor}{E} \label{e_cnt_sparse}
\end{equation} 
(assuming that the assignment is feasible).  Now, consider the more general case in which we have $E$ edges to place either in the sparse set or in the cut between the sparse set and the dense set (with $E$ being in the feasible range).  Obviously, we can place at most $E_s^{\max} = \lfloor n_s/2 \rfloor$ in the sparse set, so at least $\max\{0,E-E_s^{\max}\}$ edges must be placed within the cut.  By turns, it is obvious no more than $\min\{E,N-n_s\}$ edges can be placed within the cut.  It then follows that the total number of ways to place $E$ edges in either the sparse set or the sparse/dense interface is
\begin{equation}
C_{si}(E,n_s) = \sum_{e=\max\{0,E-E_s^{\max}\}}^{\min\{E,N-n_s\}} \binom{n_s}{e} (N-n_s)^e C_s(E-e,n_s).
\end{equation}
Since $E$ is at most order $N$, this can be computed in linear time.

\subsubsection{Multiplicity of the Dense Phase}

Now we consider internal ties within the dense phase.  The number of ways to place $E$ edges among the $N-n_s$ dense phase vertices is equivalent to the number of graphs on $N-n_s$ vertices with $E$ edges having minimum degree 2.  Although we are not aware of an exact solution for this quantity, \citet{pittel.wormald:jctA:2003} provide an asymptotic solution that can be computed in $\mathcal{O}(\log N)$ time.  As the algorithm is somewhat involved, we do not describe it in detail here, simply using $C_d(E,N-n_s)$ to denote the multiplicity of the dense phase graphs on $E$ edges.

\subsubsection{Calculating $Z$}

With the above elements, we may now describe a procedure for calculating $Z$:

\begin{algorithm}[H]
\DontPrintSemicolon
\SetKwInOut{KwInput}{Input}
\SetKwInOut{KwOutput}{Output}
\KwInput{Graph size $N$, parameter vector $\theta$}
\KwOutput{Partition function value $Z$}
Set $Z=0$\;
\For{$n_s \in 0,\ldots,N$}{ \tcp*{Walk through the number of nodes in the sparse phase}
  \For{$E_s \in 0,\ldots,n_s$}{  \tcp*{Walk through the number of edges in the sparse phase}
    \For{$E_d \in N-n_s,\ldots,\tbinom{N-n_s}{2}$}{ \tcp*{Walk through the number of edges in the dense phase}
      Set $Z:=Z+\tbinom{N}{n_s} \exp(\theta_e (E_s+E_d) + \theta_c (N-n_s)) C_{si}(E_s,n_s) C_d(E_d,N-n_s)$\;
    }
  }
}
\end{algorithm}

The time complexity of this algorithm is $\mathcal{O}(N^5)$, which while expensive is nevertheless feasible for sufficiently large $N$ to be useful.  Calculating $Z$ conditional on $n_s$ (and hence on the order parameter) requires modifying only the outer loop; this also saves us one power of $N$ in complexity.

\subsection{Phase Structure}

To study the phase transition behavior of the edge/concurrent vertex model with respect to order parameter $m$, we proceed as follows.  First, we compute $F(m,T)$ over the range $m\in [0,1]$ for various choices of $T$.  At each temperature, we identify the minima of $F_\phi$, i.e. $m^*(T) = \arg\min_{M} F_\phi(T|M)$; these are the locally stable states of the system, with respect to the order parameter.  From this we identify the \emph{critical temperature}, $T_c$, above which the sparse and dense vertex phases cannot coexist.  Examining $T/T_c$ as a function of $m^*$ allows us to map out the phase structure, and to determine the order of the associated phase transition. 

\begin{figure}
\begin{center}
\includegraphics[width=0.7\textwidth]{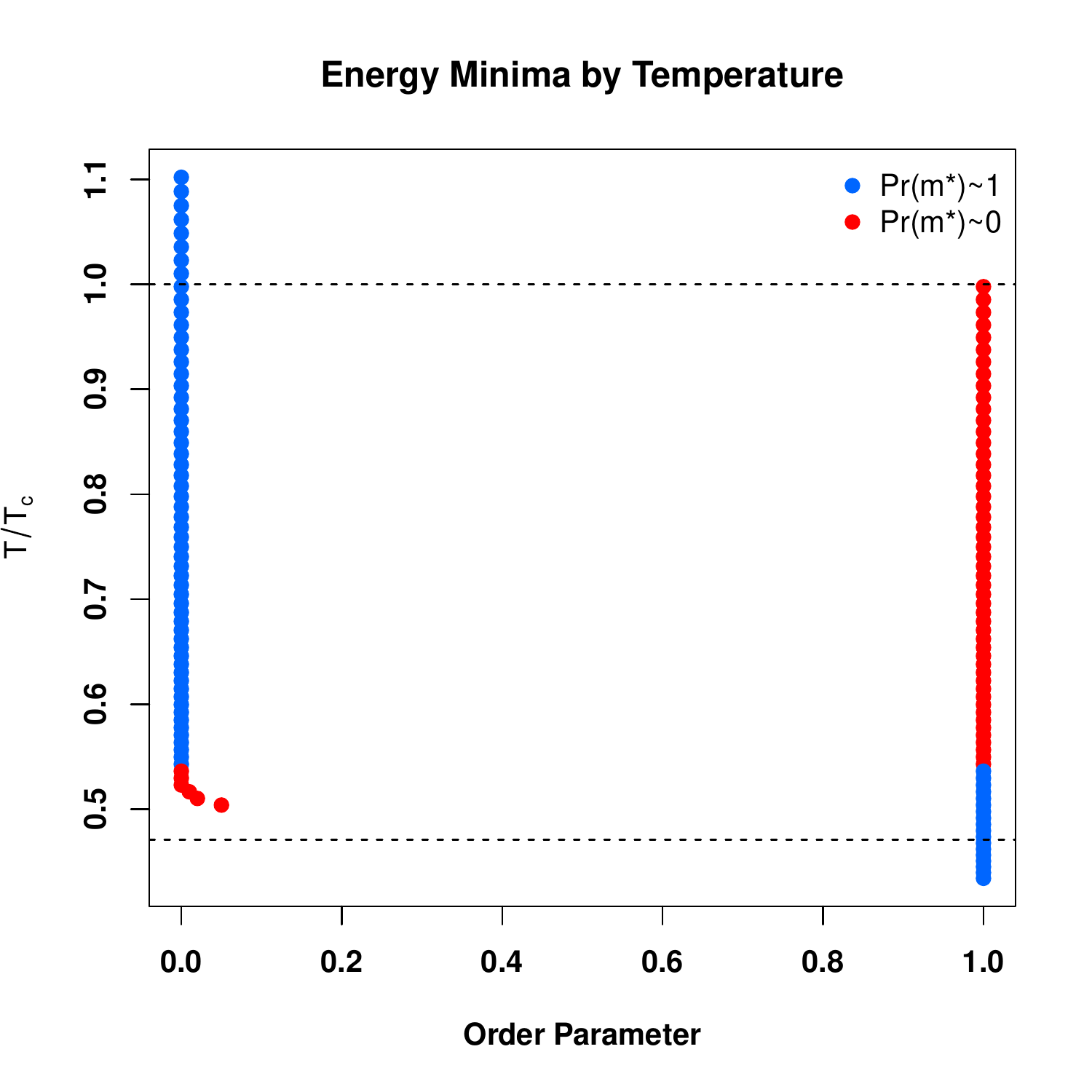}
\caption{Local free energy minima of the order parameter, as a function of temperature relative to $T_c$.  Favored states sweep from concurrency dominated above the critical temperature (top dotted line) to non-concurrent below a second threshold temperature (lower dotted line).  Where multiple minima are present, one tends to be substantially more stable (blue) than the other(s) (red).  \label{fig_minbytemp}}
\end{center}
\end{figure}

As a numerical case, we consider the specific choice $\theta_e=-1.631$, $\theta_c=-5.502$, $N=100$ examined by Krivitsky and Morris.  When expressed in terms of $\phi=(1,3.373)$, these parameters are very close to the critical temperature (numerically determined to be approximately 0.95); it is thus unsurprising that MCMC simulations using these parameter values rapidly transition to a regime in which almost all vertices are in the dense phase.  Figure~\ref{fig_minbytemp} examines this more systematically, plotting the free energy minima of the order parameter by relative temperature.  The range over which phase coexistence is possible is contained between the two horizontal dotted lines.  Above the critical temperature (upper line), the system is dominated entirely by the dense phase, with all vertices being concurrent.  Similarly, for $T/T_c$ below approximately 0.45 (lower line), almost all vertices are in the sparse phase.  At intermediate temperatures, the system is characterized by two local minima, of which one (blue) has probability approaching 1 while the other (red) is metastable (with low probability).  Stability flips in a sharp transition occurring in the $T/T_c$ range from approximately 0.55 to 0.50.  While phase coexistence is possible, in practice the system tends remain in the vicinity of the stable minimum (with rare excursions to the unstable minimum, which may or may not be observed e.g. in simulation).

\begin{figure}
\begin{center}
\includegraphics[width=0.7\textwidth]{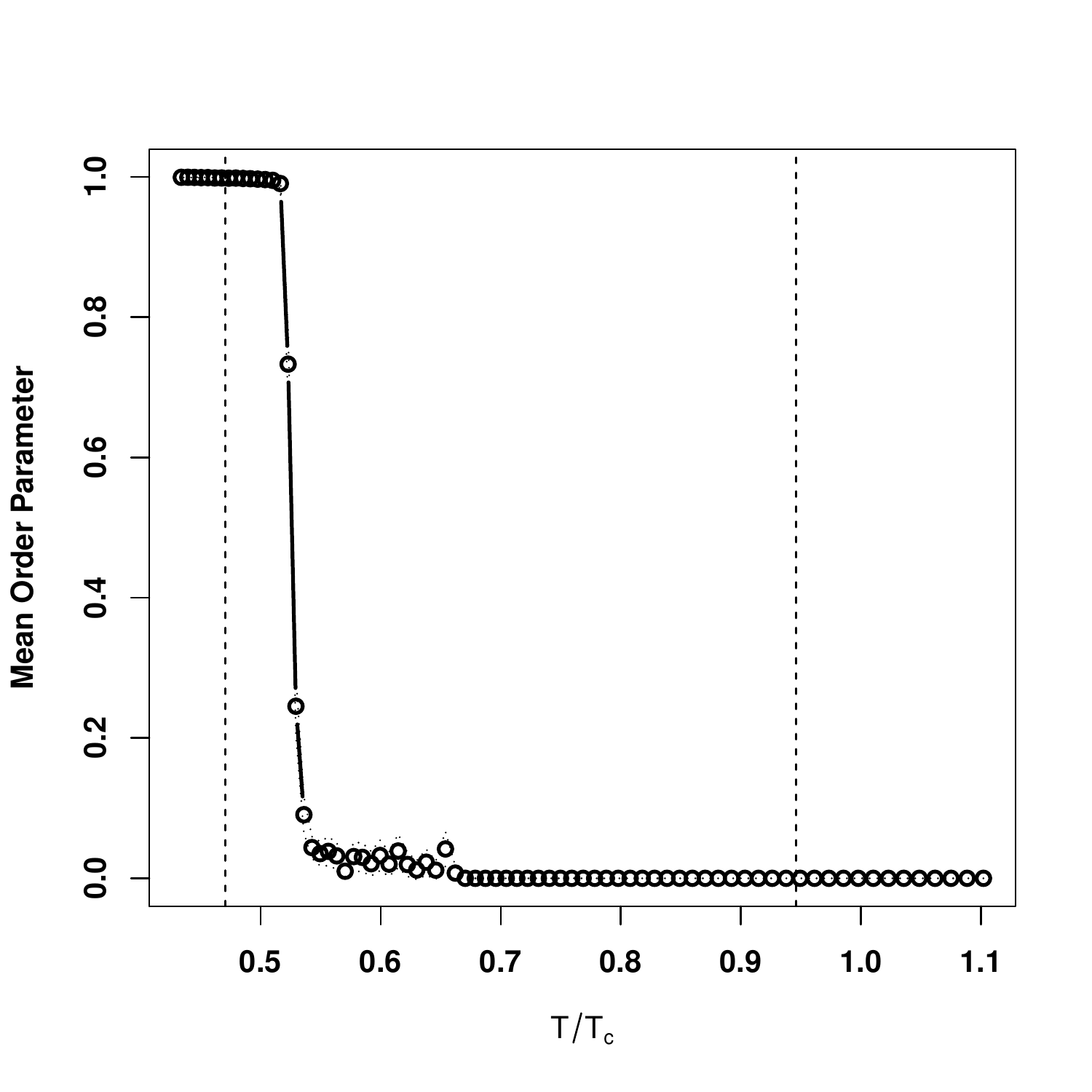}
\caption{ Mean and 95\% confidence intervals for the order parameter, as estimated from MCMC simulation. Observed mean order parameter values match behavior predicted from the free energy calculations. \label{fig_simord}}
\end{center}
\end{figure}

As a point of comparison, figure~\ref{fig_simord} provides a simulation-based estimate of the expected order parameter as a function of relative temperature.  Each point reflects the sample mean for 250 independently seeded MCMC draws from the same model as figure~\ref{fig_minbytemp} (TNT sampler, burn-in of 500,000 iterations), with seeds drawn from a Bernoulli graph distribution with expected density uniformly selected over the [0,1] interval.  (Simulation was performed using the \texttt{sna} \citep{butts:jss:2008b} and \texttt{ergm} \citep{hunter.et.al:jss:2008} libraries.)  The critical temperature and minimum coexistence temperature are shown for reference (vertical lines).  The behavior of the simulated mean order parameter closely mirrors the predictions from free energy calculations, bearing in mind that the mean involves an average over all states (instead of simply the minima themselves).  In particular, we see the stability shifts shown in figure~\ref{fig_simord}, where the dense and sparse phase swap favorabilities over a very narrow temperature range.

\begin{figure}
\begin{center}
\includegraphics[width=0.45\textwidth]{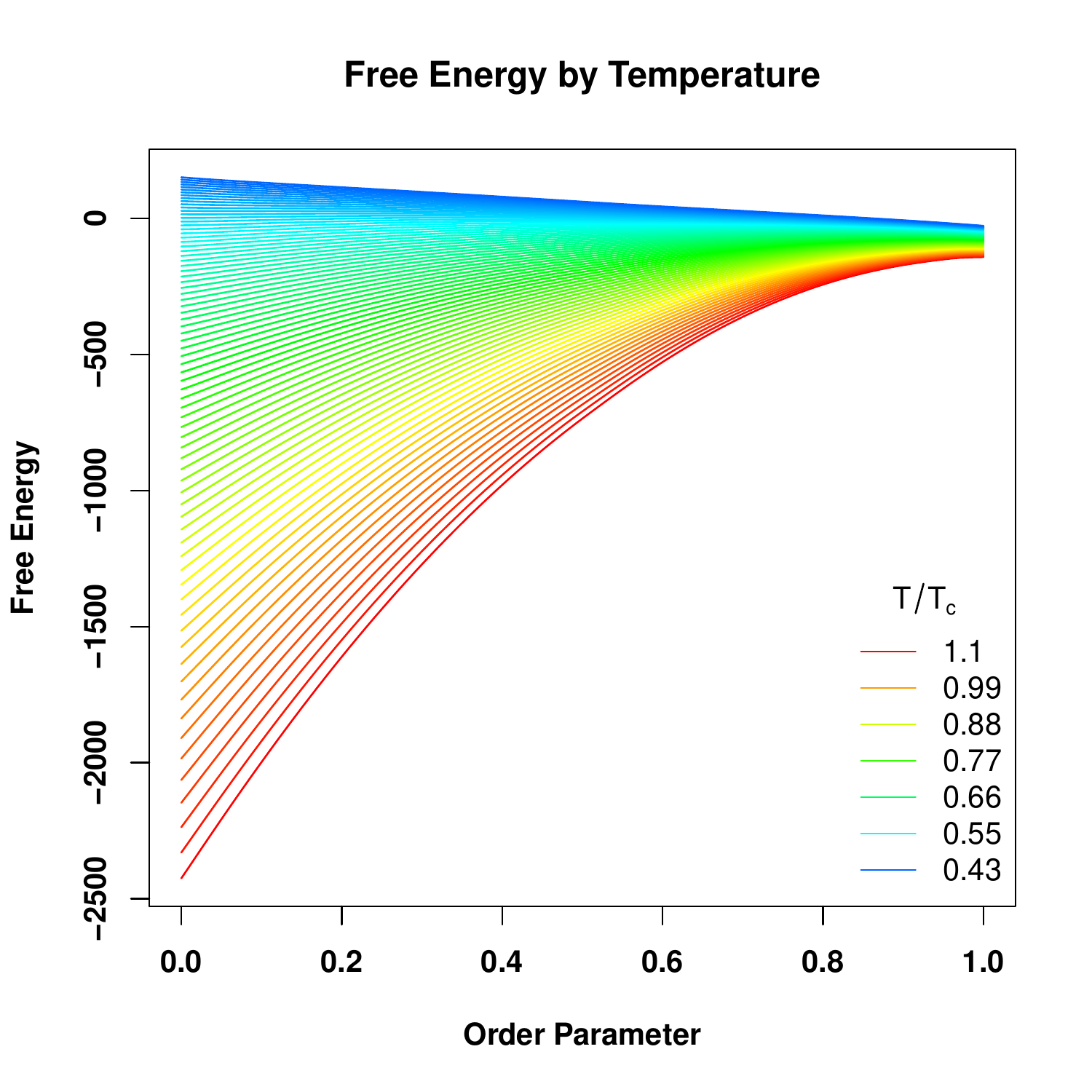}\includegraphics[width=0.45\textwidth]{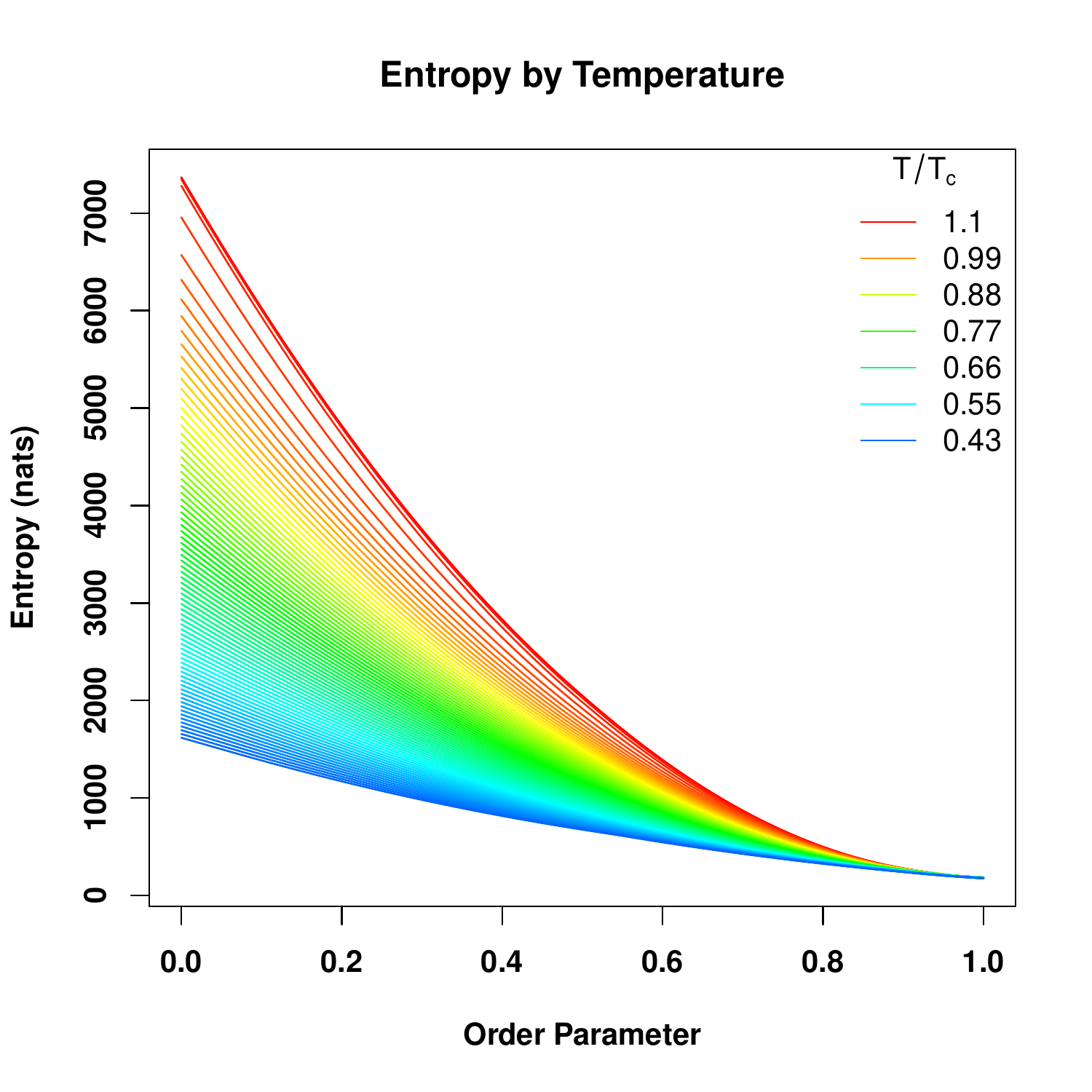}
\caption{Free energy curves (left) and entropies (right), as a function of order parameter and temperature. At high temperatures, the entropy of the concurrent phase makes it highly favorable; falling temperatures destabilize concurrency largely through reducing its entropy. \label{fig_fbytemp}}
\end{center}
\end{figure}

The role of temperature in generating the phase transition can be appreciated plotting $F$ and the entropy ($S$) as a function of the order parameter for different values of $T/T_c$.  Figure~\ref{fig_fbytemp} shows the corresponding curves for different choices of relative temperature.  Consistent with figure~\ref{fig_minbytemp}, the minima at the extreme left (high concurrency) gradually lose stability as temperature decreases, eventually leading the right-hand extreme (low concurrency) to become favored.  However, we can also see that this is effect comes about primarily via the destabilization of the dense phase at low temperature (as opposed e.g. to the stabilization of the sparse phase).  A major contributor to this phenomenon is the dramatic reduction in entropy of the dense phase seen at low temperature.  Intuitively, ``warmer'' graphs have higher mean degree, providing many more ways for unconstrained degree to tie to neighbors than those with degree $\le 1$.  At ``cooler'' temperatures, however, falling mean degree reduces this difference: when mean degrees begin to approach the dense phase lower limit (i.e., 2), the entropic advantage of the dense phase versus the sparse phase is greatly reduced.  By turns, falling temperatures also make the sparse phase more energetically favorable (both because edges are costly and because of the cost of concurrency itself).  While both effects conspire to produce a stability transition from dense to sparse, reduction in dense phase entropy is the key driving force.

\begin{figure}
\begin{center}
\includegraphics[width=0.7\textwidth]{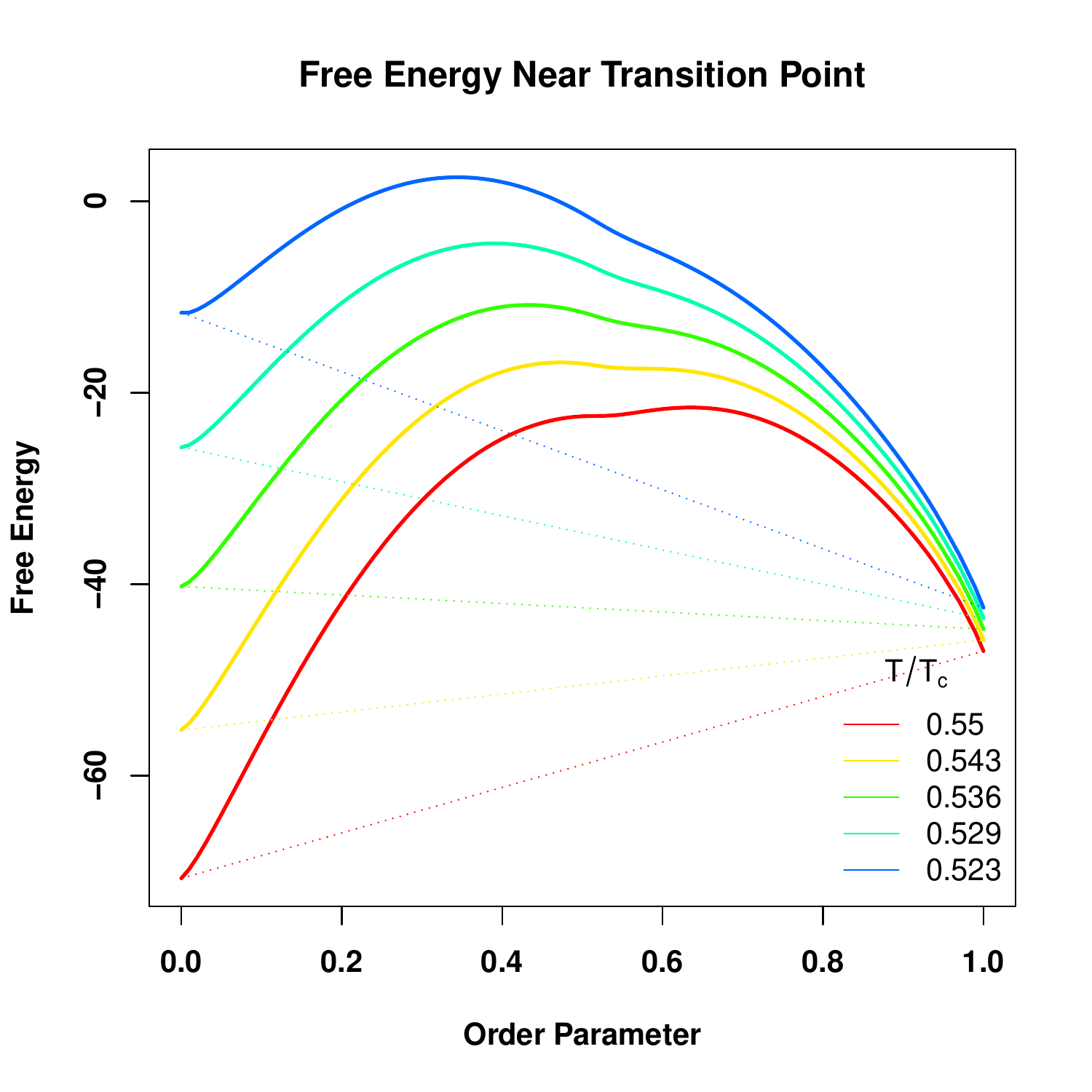}
\caption{Free energies (solid curves) near the stability transition.  The two energy minima are separated by a large barrier; note how the relative stability of the minima (dotted lines) shifts rapidly with small changes in temperature. \label{fig_fneartrans}}
\end{center}
\end{figure}

Another insight gleaned from the above is that the concurrency phase transition is first order.  This follows immediately from the presence of multiple energy minima, as seen in figure~\ref{fig_minbytemp}.  These minima are separated by a large energy barrier, whose magnitude changes (along with the relative stability of the minima themselves) with temperature.  This can be seen most clearly by examining free energy curves for temperatures near the point at which the sparse and dense phases are most equally likely.  Figure~\ref{fig_fneartrans} illustrates such a case: even where the two energy minima are relatively similar in probability, a large barrier exists that inhibits transfer between them.  Thus, the system may spend long periods dominated by one phase, before switching suddenly to the other.

\section{Weak Cooperativity in Sparse-to-Dense Trajectories}

%

The above describe some aspects of the concurrency phase transition in the edge/concurrent vertex model at the phenomenological level.  Is it possible to gain some mechanistic understanding of what drives this phenomenon (particularly in terms of trajectories that begin in the sparse phase and end in the dense phase)?  Obviously, an ERGM specification is somewhat agnostic to mechanistic detail, but nevertheless we can gain some insights by considering simple processes such as e.g. Gibbs dynamics on an ERGM of the selected form.  In this vein, consider the evolution of three vertex classes---isolates, pendants, and concurrent nodes---under the edge/concurrent vertex model in which both density and concurrency are suppressed.  The conditional probability of an $i,j$ tie being present in such a setting is one of
\begin{gather*}
p_{II/IC/CC}=\mathrm{logit}^{-1}(\theta_e),\\
p_{PI/PC}=\mathrm{logit}^{-1}(\theta_e+\theta_c), \text{ or}\\
p_{PP}=\mathrm{logit}^{-1}(\theta_e+2\theta_c),
\end{gather*}
depending on the respective class memberships of $i$ and $j$ ((P)endant, (I)solate, or (C)oncurrent).  If we begin in the ultra-sparse regime with all nodes belonging to the isolate or pendant classes, then isolates will preferentially migrate into the pendant class so long as $p_{II/IC/CC}(N-1)>1$.  When the concurrency penalty is large ($p_{PI/PC}\ll p_{II/IC/CC}$), the waiting time for a pendant to become concurrent will also be large; in particular, $p_{PP}\ll p_{PI/PC}$, and hence the merger of pendants is heavily disfavored.  Consider, however, what happens once pendant converts to a concurrent vertex: this vertex can now ``recruit'' other pendants to the concurrent class at a rate of $p_{PI/PC} \gg p_{PP}$, each of whom can in turn recruit others at that same rate.  By turns, a concurrent vertex can form additional ties with other concurrent vertices at pairwise rate $p_{II/IC/CC}$, and so long as $p_{II/IC/CC}(N-1)$ is substantially larger than 1, such vertices will rarely lose enough ties to drop back into the pendant class.  The net effect of this process is that vertices that escape into the dense phase are ``lost'' to the sparse phase, and each such vertex accelerates the capture of more sparse phase vertices (by creating more opportunities for pendants to form additional ties at rate $p_{PI/PC} \gg p_{PP}$).  Unless the respective rates are very precisely balanced (which occurs only over an extremely narrow temperature range), the net result is an extremely rapid and locally irreversible transfer of vertices from the sparse phase to the dense phase once the first concurrent vertices arise.

\begin{figure}
\begin{center}
\includegraphics[width=0.7\textwidth]{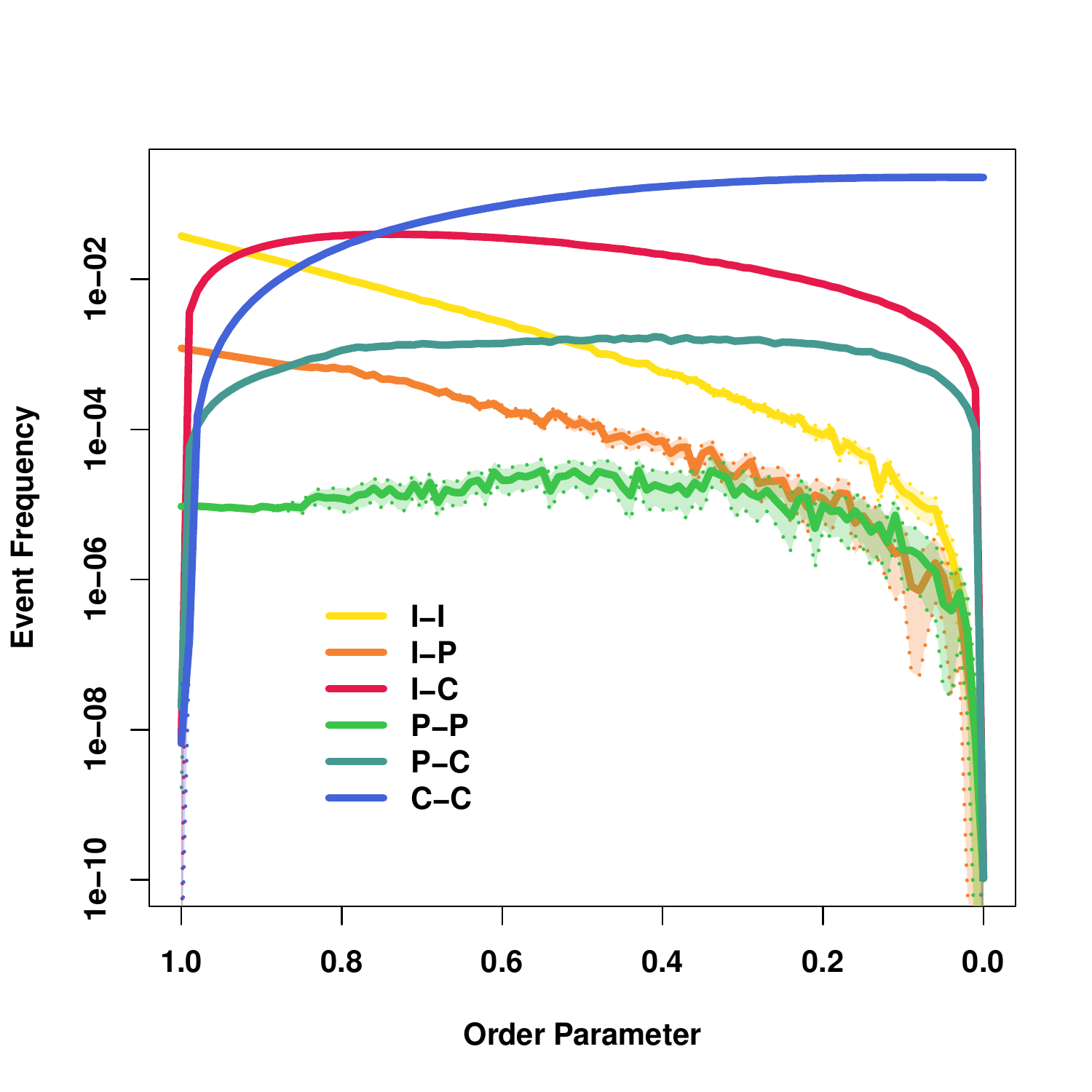}
\caption{Edge formation event rates under Metropolis dynamics, as a function of order parameter (posterior mean estimates; shaded areas indicate 95\% posterior intervals, where wide enough to be visible). Conversion of isolates to pendants is facilitated by concurrent vertices (I-C tie formation), while concurrent vertices quickly accumulate enough ties to resist conversion to pendants (C-C tie formation).  Concurrent vertices are then able to irreversibly convert pendants to the concurrent class (P-C tie formation).  By contrast, P-P ties are always highly disfavored.  \label{fig_formevent}}
\end{center}
\end{figure}

Examination of mean edge formation event rates under Metropolis dynamics as a function of the order parameter corroborates this interpretation (Figure~\ref{fig_formevent}).  Using rejection sampling, we sample 1,000 MCMC trajectories making a full transition from the sparse to the dense phase within $5\times 10^7$ iterations, sampling every fifth transition in each selected trajectory; tabulating the event types of the sampled MCMC transitions against the order parameter allows us to estimate the mean rate for each type of edge formation event in each step of the phase transition.  (Simulations performed using \texttt{ergm} \citep{hunter.et.al:jss:2008} with uniform random change proposals.)  Isolate-isolate tie formation initially dominates, but is rapidly overtaken by isolate-concurrent tie formation once even a small number of concurrent vertices is present.  At the same time, rapid tie formation among concurrent vertices prevents them from returning to the sparse phase.  While recruitment of pendants into the concurrent class is slower than recruitment into the pendant class, concurrent-pendant ties eventually drain the latter class into the dense phase.  Since pendant-pendant tie formation is exceedingly rare, recruitment is driven by interaction with concurrent nodes.

This phenomenon can be understood as a form of weak cooperativity, where previously recruited concurrent vertices accelerate the rate at which new vertices are recruited.  Unlike e.g. the process of runaway triangle formation in the edge/triangle or edge/$k$-star models, this rate enhancement is bounded by a factor of approximately $\exp(\theta_c)$, resulting in a slower transition; moreover, in this case the dense phase is entropically favored, whereas the sparse phase is favored under these other models.  This mechanism is thus distinct, though it also has a cooperative character. 

\section{Discussion and Conclusion}

Despite its simplicity, the edge/concurrent vertex model can exhibit interesting phase behavior.  As shown here, it displays a first order phase transition with respect to an order parameter associated with the fraction of non-concurrent vertices, transiting from a high-temperature, Bernoulli-like regime in which essentially all vertices are concurrent to a low-temperature, ultra-sparse regime in which almost all vertices are pendants or isolates.  Near the critical temperature, trajectories originating in the ultra-sparse regime transit to the dense regime by a weakly cooperative process, in which vertices recruited to the concurrent phase facilitate the transition of non-concurrent vertices to concurrency. As the concurrent phase is entropically favored, concurrent vertices do not readily convert back to isolates, leading to a ``ratchet'' effect that eventually consumes the entire system.

Given its use as a model for sexual contact networks, it is perhaps appropriate to consider whether this behavior of the edge/concurrent vertex model is a realistic depiction of social life.  For this application, the behavior described here would seem to be less than ideal: although lurid tales of hedonists who draw others into lives of promiscuity have been a recurrent feature of urban legends and morality tales, reliable empirical accounts of such phenomena are difficult to come by.  The weak cooperativity that drives the concurrency phase transition is thus suspect (at least for sexual contact networks), and may be undesirable.

Since the source of cooperativity in the edge/concurrent vertex model stems from the fact that only the second tie incident upon a given vertex is penalized, an obvious alternative is to penalize \emph{each} tie after the first.  The edge/concurrent tie model does precisely this, and may be a better choice.  Alternately, it is plausible in many settings that each additional tie is more difficult to form and sustain than the one before it (e.g., because of the increasing constraints imposed by each partnership on time and resources, in addition to the increased opportunities for conflict among partners incurred with higher degree).  In this scenario, an edge/2-star model could be plausible.  Although the edge/2-star model is often seen as too degeneracy-prone to be useful, these conclusions are generally based on cases in which the 2-star parameter is positive (usually leading to density explosion).  A \emph{negative} 2-star term leads to very different behavior, however, and may be worthy of further investigation.  

The richness of models like that studied here is a reminder that even simple models of social systems can have complex emergent behavior.  In this light, is somewhat ironic that phase transitions were objects of such fascination in the 1970s and 1980s that contemporary researchers expressed concern that models were being engineered to produce them for their own sake, without concern for empirical plausibility \citep{fararo:bs:1978, fararo:bs:1984}.  In the context of modern network models, we find phenomena such as phase transitions arising unbidden and often unrecognized from very basic social mechanisms.  Understanding this behavior may require theoretical tools that go beyond those traditionally used to study the behavior of statistical network models.  For instance, it is interesting to observe in the present case that the ultra-sparse phase exhibits a type of invariance (``symmetry,'' broadly understood) that the dense phase does not.  Specifically, both isolates and 2-cliques (the only structures possible in pure sparse phase) have the property that certain pairs of vertices can be permuted without altering the structure; put another way, vertices in the sparse phase occupy one of only two automorphic equivalence classes \citep{everett:sn:1985}, leading to a large automorphism group with a very particular structure.  By contrast, the concurrent phase behaves roughly like a typical Bernoulli graph, which tends to have few if any nontrivial automorphisms \citep[see e.g.][Theorem 2]{erdos.renyi:amh:1963}.  This difference in symmetry is a major driver in the entropic cost of the sparse phase, and it is plausible that insights into the phase transition behavior of the edge/concurrent vertex model can be obtained directly from an analysis of the phases' automorphism groups.  Though symmetry has played a core role in the development of classical network analysis (from structural equivalence \cite{lorrain.white:jms:1971} and role algebras \cite{white:bk:1963,boyd:jmp:1969,pattison:bk:1993} to generalized blockmodeling \cite{doreian.et.al:bk:2005}), it has been less prominent in our attempts to understand the behavior of complex random graph models.  Further studies of ``Drosophila'' models, where such properties are relatively tractable, have the potential to yield considerable insight into the factors that drive social structure, and to inform the design of more complex models for empirical use.

\bibliography{ctb}


\end{document}